\documentstyle[12pt]{article}
\newcommand{\braket}[1]{\langle #1 \rangle}
\newcommand{\graket}[1]{\langle #1 \rangle_{{\rm g}}}

\begin{document}
\begin{center}
\par
\noindent
{\LARGE {\bf Ground-State Phase Diagram}}
\par
\noindent
{\LARGE {\bf of the Two-Dimensional}}
\par
\noindent
{\LARGE {\bf Quantum Heisenberg Mattis Model\footnote{J.\ Phys.\ Soc.\ Jpn.\
{\bf 65} (1996) No.\ 2 in press.}}}
\bigskip
\bigskip
\par
\noindent
{\Large
Yoshihiko {\sc Nonomura}\footnote{Present address:
Department of Physics, University of Tokyo,
Hongo 7-3-1, Bunkyo-ku, Tokyo 113, Japan.}
}
\bigskip
\par
\noindent
{\large
{\it Department of Physics, Tokyo Institute of Technology,}\\
{\it Oh-okayama 2-12-1, Meguro-ku, Tokyo 152, Japan}
}
\medskip
\par
\noindent
(Received: March 15, 1995)
\bigskip
\bigskip
\par
\noindent
{\bf Abstract}
\end{center}
The two-dimensional $S=1/2$ asymmetric Heisenberg Mattis model is
investigated with the exact diagonalization of finite clusters. The
N\'eel order parameter and the spin glass order parameter can be smoothly
extrapolated to the thermodynamic limit in the antiferromagnetic region,
as in the pure Heisenberg antiferromagnet. The critical concentration of
the N\'eel phase is consistent with that of the two-dimensional Ising
Mattis model, and the spin glass order parameter increases monotonously
as the ferro-bond concentration increases. These facts suggest that
quantum fluctuation does not play an essential role in
two-dimensional non-frustrated random spin systems.
\bigskip
\section{Introduction}
In recent studies on spin systems, effects of quantum fluctuation and
randomness have been investigated from various aspects. In considering
recent development of computer facilities, it is high time to study
the coexistence of these two nontrivial effects. In fact, numerous
studies have already been made for the infinite-range Ising model
in a transverse field\cite{sktris} and the infinite-range quantum
XY model.\cite{skqxy} Quite recently, analytic and numerical studies
of the short-range random Heisenberg model began to be attempted in
one-\cite{dsf1,dsf2,ngft} and two-dimensional\cite{2dpmj,2ddim} systems.
However, all these works are motivated by the question how ground-state
properties of quantum spin systems are affected by random perturbations,
and the behavior of characteristic quantities in random spin systems
such as the spin glass (SG) order has not been considered until present.

As a first step to quantum SG problems, we investigate the
two-dimensional $S=1/2$ asymmetric Heisenberg Mattis model
on a square lattice described by the following Hamiltonian,
\begin{equation}
  \label{qmatham}
  {\cal H}=-\sum_{\braket{ij}}J_{ij}\vec{S}_{i}\cdot\vec{S}_{j}\ ,
\end{equation}
where $\braket{ij}$ denotes all the nearest-neighbor pairs, with
\begin{eqnarray}
  \label{asm1}
  P(J_{ij};p)&=&p\delta(J_{ij}-J)+(1-p)\delta(J_{ij}+J)\ ,\\*
  J_{ij}     &=&J\tau_{i}\tau_{j}\ ,\ \ \tau_{i}=\pm 1\ .
  \label{asm2}
\end{eqnarray}
Since frustration does not exist\cite{mattis} in the present model,
the coexistence effect of quantum fluctuation and randomness
is expected to be displayed clearly. As is well-known,
the free energy of the following Ising Mattis model,
\begin{equation}
  \label{ismatham}
  {\cal H}=-\sum_{\braket{ij}}J_{ij}S_{i}^{z}S_{j}^{z}\ ,
  \ \ {\textstyle S=\frac{1}{2}}\ ,
\end{equation}
with the conditions (\ref{asm1}) and (\ref{asm2}), is
equivalent\cite{mattis} to that of the corresponding
pure Ising model through the following gauge transformation
of the spin variable $\{S_{i}^{z}\}$,
\begin{equation}
  \label{istr}
  S_{i}^{z}\rightarrow S_{i}^{z}\tau_{i}\ .
\end{equation}
On the other hand, such equivalence does not hold in the quantum
Heisenberg Mattis model, and ground-state properties of this model
are nontrivial. In the present paper, we investigate the behavior
of the N\'eel order parameter and the SG order parameter.

In \S 2, properties of the asymmetric Ising Mattis model
are briefly reviewed, following Ozeki's paper.\cite{asmat} His
argument is extended to the asymmetric quantum XY Mattis model.
In \S 3, the finite-size correction of the SG order parameters
is investigated by means of the exact diagonalization of the
two-dimensional $S=1/2$ XY model. In \S 4, ground-state
properties of the two-dimensional $S=1/2$ Heisenberg Mattis model are
investigated similarly. The critical concentration $p_{{\rm c}}^{{\rm AF}}$
of the N\'eel phase and the $p$-dependence of the SG order parameter are
calculated. In \S 5, these descriptions are summarized.
\section{Review of the Asymmetric Mattis Model}
Even in the classical Ising-spin case, the Mattis model\cite{mattis}
shows nontrivial phase transitions\cite{asmat} when asymmetric bond
distribution ($p\neq 0.5$) is considered. Namely, the phase diagram of
the Ising Mattis model (\ref{ismatham}) is shown in Fig.\ \ref{asmatfig},
and this phase diagram can be obtained {\it analytically} from that of the
corresponding pure Ising model. This mechanism can be understood through
the behavior of correlation functions. Following Ozeki's paper,\cite{asmat}
let us consider the ferromagnetic correlation function,
\begin{equation}
  g(r;p,T)\equiv [\braket{S_{0}^{z}S_{r}^{z}}]_{{\rm r}}\ ,
\end{equation}
and the SG correlation function,
\begin{equation}
  \tilde{g}(r;p,T)\equiv [\braket{S_{0}^{z}S_{r}^{z}}^{2}]_{{\rm r}}\ .
\end{equation}
Here $\braket{\cdots}$ denotes the thermal average,
and $\left[\cdots\right]_{{\rm r}}$ represents the
random average based on the conditions (\ref{asm1}) and
(\ref{asm2}). After the gauge transformation (\ref{istr}),
these two correlation functions can be expressed as
\begin{eqnarray}
  g(r;p,T)        &=&\braket{S_{0}^{z}S_{r}^{z}}_{{\rm pure}}\times
                     \left[\tau_{0}\tau_{r}\right]_{{\rm r}}\ ,\\*
  \tilde{g}(r;p,T)&=&\braket{S_{0}^{z}S_{r}^{z}}_{{\rm pure}}^{2}\times
                     \left[(\tau_{0}\tau_{r}\right)^{2}]_{{\rm r}}\nonumber\\*
                  &=&\braket{S_{0}^{z}S_{r}^{z}}_{{\rm pure}}^{2}\ ,
\end{eqnarray}
where $\braket{\cdots}_{{\rm pure}}$ denotes the thermal average
in the pure Ising model ($p=1$ case of the model (\ref{ismatham})).
Then, the structure of the phase diagram given in Fig.\ \ref{asmatfig}
can be explained clearly. The para-ferro (or the para-Mattis SG) phase
transition originates from the singularity of the correlation
function $\braket{S_{0}^{z}S_{r}^{z}}_{{\rm pure}}$,
and the ferro-Mattis SG phase transition originates
from that of $\left[\tau_{0}\tau_{r}\right]_{{\rm r}}$.
Since the latter correlation function is independent
of temperature, the ferro-Mattis SG phase transition
occurs even in the ground state. Moreover, this random
average can be related\cite{asmat} with the thermal average
$\braket{\tau_{0}^{z}\tau_{r}^{z}}$ defined in the following Ising model,
\begin{eqnarray}
  \label{altis1}
  {\cal H}_{{\rm F}}\{\tau\}&=&-\sum_{\braket{ij}}\tau_{i}\tau_{j}
  \ \ \ \mbox{for}\ \ p\geq 1/2\ ,\\*
  {\cal H}_{{\rm AF}}\{\tau\}&=&+\sum_{\braket{ij}}\tau_{i}\tau_{j}
  \ \ \ \mbox{for}\ \ p\leq 1/2\ .
  \label{altis2}
\end{eqnarray}
In this argument, the difference between the micro-canonical ensemble
(the original bond distribution based on the conditions (\ref{asm1}) and
(\ref{asm2})) and the canonical ensemble (with respect to the Ising models
(\ref{altis1}) and (\ref{altis2})). Then, the critical concentration
$p_{{\rm c}}^{{\rm F/AF}}$ of this model is related\cite{asmat}
with the critical temperature $T_{{\rm c}}$ of the Ising models
(\ref{altis1}) and (\ref{altis2}), as
\begin{eqnarray}
  e_{{\rm F}}(T_{{\rm c}}) &=&1-2p_{{\rm c}}^{{\rm F}}\ ,\\*
  e_{{\rm AF}}(T_{{\rm c}})&=&2p_{{\rm c}}^{{\rm AF}}-1\ ,
\end{eqnarray}
where $e_{{\rm F}}$ and $e_{{\rm AF}}$ represent the energies per bond of
${\cal H}_{{\rm F}}\{\tau\}$ and ${\cal H}_{{\rm AF}}\{\tau\}$, respectively.
Therefore, these two phase transitions at $T=T_{{\rm c}}$ or
$p=p_{{\rm c}}^{{\rm F/AF}}$ belong to the Ising universality
class. Especially, in the two-dimensional Mattis model on a
square lattice, the critical temperature and the critical
concentration are given\cite{asmat} by $T_{{\rm c}}=2.269\cdots$ and
$p_{{\rm c}}^{{\rm F}}=1-p_{{\rm c}}^{{\rm AF}}=0.853\cdots$, respectively.

Ozeki's argument is valid even in quantum spin systems, as long as
they can be transformed to pure systems through similar gauge
transformations.\cite{matxy} Namely, the quantum XY Mattis
model described by the following Hamiltonian,
\begin{equation}
  \label{xymatham}
  {\cal H}=-\sum_{\braket{ij}}J_{ij}(S_{i}^{x}S_{j}^{x}+S_{i}^{y}S_{j}^{y})\ ,
\end{equation}
with the conditions (\ref{asm1}) and (\ref{asm2}), can be
treated similarly through the following transformations,
\begin{eqnarray}
  \label{xytr1st}
  S_{i}^{x}&\rightarrow&S_{i}^{x}\tau_{i}\ ,\\*
  S_{i}^{y}&\rightarrow&S_{i}^{y}\tau_{i}\ ,\\*
  S_{i}^{z}&\rightarrow&S_{i}^{z}\ .
  \label{xytr3rd}
\end{eqnarray}
Since the existence of the long-range order has already been proved in the
pure two-dimensional $S=1/2$ XY model at the ground state,\cite{xylro}
the ground-state phase diagram of the model (\ref{xymatham}) should
be equivalent to that of the two-dimensional Ising Mattis model:
The critical concentrations are given by
$p_{{\rm c}}^{{\rm F}}=1-p_{{\rm c}}^{{\rm AF}}=0.853\cdots$,
and the value of the SG order parameter is independent of $p$.
\section{Finite-Size Correction of the Spin-Glass Order Parameter}
The spin-wave theory tells that the asymptotic form of the two-point
correlation function of the two-dimensional antiferromagnetic
Heisenberg model ($p=0$ case of the model (\ref{qmatham}))
at the ground state is given\cite{swcorr,modsw} by
\begin{equation}
  \label{neelcorr}
  \graket{\vec{S}_{i}\cdot\vec{S}_{j}}
  \sim (-1)^{|i-j|}\left(m_{0}+\frac{1}{\sqrt{2}\pi r_{ij}}\right)^{2}\ ,
\end{equation}
with $r_{ij}\equiv\left|\vec{r}_{i}-\vec{r}_{j}\right|$, and
$\graket{\cdots}$ represents the ground-state average. Thus,
the N\'eel order parameter $m_{{\rm st}}(N)$ defined in
\begin{equation}
  \label{neelop}
  m_{{\rm st}}^{2}(N)\equiv\frac{1}{N^{2}}\sum_{i,j}(-1)^{|i-j|}
    \left[\graket{\vec{S}_{i}\cdot\vec{S}_{j}}\right]_{{\rm r}}\ ,
\end{equation}
and the SG order parameter $m_{{\rm sg}}(N)$ defined in
\begin{equation}
  \label{sgop}
  m_{{\rm sg}}^{2}(N)\equiv\frac{1}{N^{2}}\sum_{i,j}
    \left[\graket{\vec{S}_{i}\cdot\vec{S}_{j}}^{2}\right]_{{\rm r}}\ ,
\end{equation}
are expected to be scaled as
\begin{eqnarray}
  \label{magsc}
  m_{{\rm st}}^{2}(N)&\sim&m_{0}^{2}
                           +{\rm const.}\times N^{-1/2}+O(N^{-1})\ ,\\*
  m_{{\rm sg}}^{2}(N)&\sim&m_{0}^{4}
                           +{\rm const.}\times N^{-1/2}+O(N^{-1})\ .
  \label{sgsc}
\end{eqnarray}
These formulas mean $m_{{\rm st}}(\infty)=m_{0}$ and
$m_{{\rm sg}}(\infty)=m_{0}^{2}$, and the scaling form (\ref{magsc})
was confirmed by quantum Monte Carlo simulations\cite{2dheiqmc,2dxyqmc}
of this model. Although these scaling forms are not trivial when
randomness is introduced, essential behavior is expected to
be similar in the N\'eel phase, as in the diluted quantum
antiferromagnet.\cite{2ddil1,2ddil2} On the other hand,
the N\'eel order does not exist in the Mattis SG phase
($p\approx 0.5$), and the asymptotic form of the
two-point correlation function is supposed to be
\begin{equation}
  \graket{\vec{S}_{i}\cdot\vec{S}_{j}}
  \sim (-1)^{|i-j|}\exp\left(-\frac{\xi}{r_{ij}}\right)\ .
\end{equation}
Then, the scaling forms of the order parameters are given by
\begin{eqnarray}
  \label{dismagsc}
  m_{{\rm st}}^{2}(N)&\sim&{\rm const.}\times N^{-1}+O(N^{-2})\ ,\\*
  m_{{\rm sg}}^{2}(N)&\sim&{\rm const.}\times N^{-1}+O(N^{-2})\ .
\end{eqnarray}
Namely, the existence of the SG order parameter in the Mattis SG phase
cannot be derived from the asymptotic form of the two-point correlation
function. Therefore, the scaling form (\ref{sgsc}) obtained from
(\ref{neelcorr}) is also questionable. In fact, when $m_{{\rm sg}}^{2}(N)$
of the antiferromagnetic Heisenberg model is plotted versus $N^{-1/2}$
as predicted in (\ref{sgsc}) (Fig.\ \ref{sghffig}), this order parameter
does not seem to exist in the thermodynamic limit, which is not
consistent with the existence of the N\'eel order parameter.

Then, the size dependence of the SG order parameter can only be argued
numerically at present. For this purpose, the two-dimensional $S=1/2$ XY
model is considered. According to spin-wave analyses and quantum Monte
Carlo simulations of this model,\cite{2dxyqmc} the size dependence
of the N\'eel order parameter is equivalent to that of the
two-dimensional $S=1/2$ antiferromagnetic Heisenberg model. This fact
is due to the similarity of the structure of low-energy excitations in
these two models, and the size dependence of the SG order parameter of
the XY model is also expected to be as such. Moreover, even if the
Mattis-type randomness is introduced, the value of the SG order
parameter does not vary in this model, as explained in the previous
section. Since the $x,y$-axes and the $z$-axis are not equivalent in the
XY model, we should consider two different SG order parameters defined in
\begin{equation}
  \label{sgopx}
  (m_{{\rm sg}}^{x}(N))^{2}\equiv\frac{1}{N^{2}}\sum_{i,j}
    \left[\graket{S_{i}^{x}S_{j}^{x}}^{2}\right]_{{\rm r}}
\end{equation}
and
\begin{equation}
  \label{sgopz}
  (m_{{\rm sg}}^{z}(N))^{2}\equiv\frac{1}{N^{2}}\sum_{i,j}
    \left[\graket{S_{i}^{z}S_{j}^{z}}^{2}\right]_{{\rm r}}\ .
\end{equation}
We diagonalize Oitmaa-Betts-type clusters\cite{ob} up to the $20$-spin one,
and calculate these order parameters. The quantity $m_{{\rm sg}}^{z}$
should vanish in the thermodynamic limit, and $m_{{\rm sg}}^{x}$ should
remain finite. In addition, the size dependence of these two order
parameters is expected to be similar. All these conditions are
satisfied when these order parameters are plotted versus $N^{-1}$,
as shown in Figs.\ \ref{xysgzfig} and \ref{xysgxfig}. It is true that we
cannot rule out the possibility of $\sim N^{-\theta}$ ($\theta\sim 1$,
but not $1$) dependence only from these numerical data, but such a
fractional exponent can only appear when the system is critical.
Thus, we assume the scaling form of the SG order parameter as
\begin{equation}
  \label{dissgsc}
  m_{{\rm sg}}^{2}(N)\sim m_{{\rm sg}}^{2}(\infty)
                          +{\rm const.}\times N^{-1}+O(N^{-2})\ .
\end{equation}
\section{Numerical Results of the Heisenberg Mattis Model}
As in the previous section, we diagonalize Oitmaa-Betts-type clusters
up to the $20$-spin one. Note that this cluster size is not so ``small"
in the sense of numerical calculation, because spatial symmetries cannot
be used in the Hamiltonian-matrix diagonalization in random spin systems.
In compensation for such limitation of cluster sizes, we average all
the bond configurations. Since spatial symmetries can be applied to
this summation, such calculations are within the reach of recent
super computers, though still tedious. Then, the present results
are ``exact", not only in the sense of Hamiltonian diagonalization,
but also in the sense of random averaging in finite clusters.

In an $N$-spin Oitmaa-Betts-type cluster, the number of bonds are $2N$, and
the number of ferromagnetic bonds which satisfy the conditions (\ref{asm1})
and (\ref{asm2}) are limited to $N_{{\rm F}}=0,4,6,\cdots,2N-4,2N$.
Then, in order to calculate a physical quantity $Q$ for an arbitrary
concentration $p$, we should extrapolate the values of $Q$ at
$\tilde{p}\equiv N_{{\rm F}}/2N$. In the present paper,
we sum up all the bond configurations on the basis of the
canonical distribution\cite{asmat} of the corresponding $N$-spin
Ising model. Namely, using the ``temperature" $T(p)$ determined by
\begin{eqnarray}
  e_{{\rm F}}(T(p)) &=&1-2p\ \ \ \mbox{for}\ \ p\geq 1/2\ ,\\*
  e_{{\rm AF}}(T(p))&=&2p-1\ \ \ \mbox{for}\ \ p\leq 1/2\ ,
\end{eqnarray}
we have
\begin{equation}
  Q(p)=\frac{\sum_{\tilde{p}}w(\tilde{p})\tilde{Q}(\tilde{p})
             e^{-E_{g}(\tilde{p})/k_{{\rm B}}T(p)}}
            {\sum_{\tilde{p}}w(\tilde{p})
             e^{-E_{g}(\tilde{p})/k_{{\rm B}}T(p)}},
\end{equation}
where $w(\tilde{p})$ denotes the number of bond configurations in which
the ferro-bond concentration is equal to $\tilde{p}$, $\tilde{Q}(\tilde{p})$
represents the partial random average restricted to the samples in which the
ferro-bond concentration is equal to $\tilde{p}$, and $E_{g}(\tilde{p})$ is
given by
\begin{eqnarray}
  E_{g}(\tilde{p})&=&2N(1-2\tilde{p})\ \ \ \mbox{for}\ \ p\geq 1/2\ ,\\*
  E_{g}(\tilde{p})&=&2N(2\tilde{p}-1)\ \ \ \mbox{for}\ \ p\leq 1/2\ .
\end{eqnarray}

At first, we estimate the N\'eel-Mattis SG phase boundary
by means of the least-squares fitting of the N\'eel order parameter
$m_{{\rm st}}^{2}$. In Fig.\ \ref{magfitfig}, $m_{{\rm st}}^{2}$ for
the $8,10,16,18,20$-spin clusters are plotted versus $N^{-1/2}$ for
$p=0.0,0.1,0.15,0.2$. This figure shows that $m_{{\rm st}}^{2}$ vanishes
at $p\simeq 0.15$. Then, we vary the value of $p$ more precisely, and obtain
the following estimate of the critical concentration of this phase boundary,
\begin{equation}
  \label{pcval}
  p_{{\rm c}}^{{\rm AF}}=0.140\pm 0.016\ .
\end{equation}
This error bar means that the estimate of $m_{{\rm st}}$ does
not coincide with zero for $p\leq 0.124$ and $p\ge 0.156$,
within the standard deviation in the fitting.
This estimate is consistent with that of
the Ising Mattis model,\cite{asmat} $p_{{\rm c}}^{{\rm AF}}\simeq 0.147$.
Although clusters are still small, the fitting curves are linear enough,
as in the pure quantum antiferromagnet. The reason of such a good scaling
behavior may be that wavefunctions are not localized in antiferromagnetic
random spin systems owing to quantum fluctuation. On the other hand,
similar analysis is not possible for the ferromagnetic order parameter,
because quantum fluctuation is small in the ferromagnetic region.

For $p>p_{{\rm c}}^{{\rm AF}}$, the extrapolated value of $m_{{\rm st}}^{2}$
is negative, which simply means that the scaling form (\ref{magsc}) based
on the existence of the N\'eel order is not correct. The expected scaling
form in this region is (\ref{dismagsc}), and this form is satisfied in
the region $p\approx 0.5$. The linearity of the fitting curve just above
the critical concentration is expected to be a finite-size effect. Moreover,
the size dependence of the N\'eel order parameter can be fractional at
$p=p_{{\rm c}}^{{\rm AF}}$ as $m_{{\rm st}}^{2}(N)\sim N^{-\theta}$,
with $1/2<\theta<1$. These facts show that the estimate (\ref{pcval})
may be underestimate. However, as long as $p<p_{{\rm c}}^{{\rm AF}}$,
the size dependence (\ref{magsc}) holds no matter how small $m_{{\rm st}}$
is, and the present naive fitting can be justified practically. In fact,
the critical concentration of the diluted antiferromagnetic Heisenberg
model was analyzed similarly.\cite{2ddil2} Furthermore, it is quite
improbable that the critical concentration is increased by quantum
fluctuation. Therefore, the good coincidence of the estimate (\ref{pcval})
and the classical value $p_{{\rm c}}^{{\rm AF}}\simeq 0.147$ can be
regarded as the evidence of the smallness of such finite-size effects.

\begin{figure}[h]
\caption{}
\label{sgfitfig}
\end{figure}
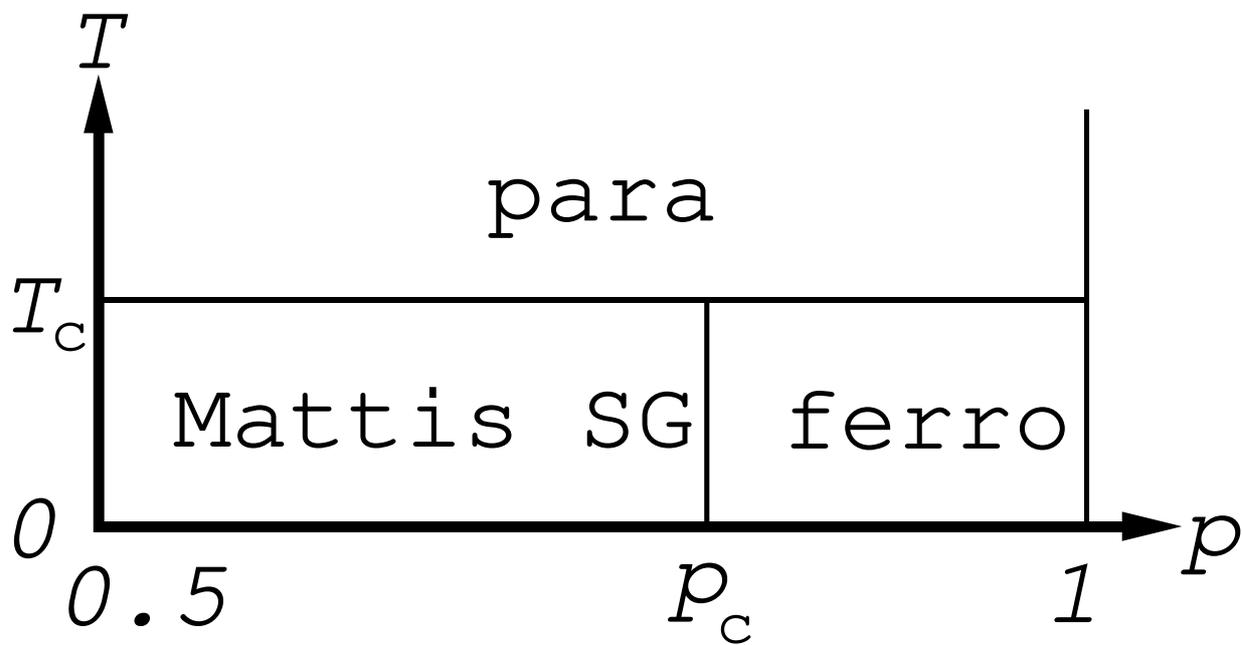
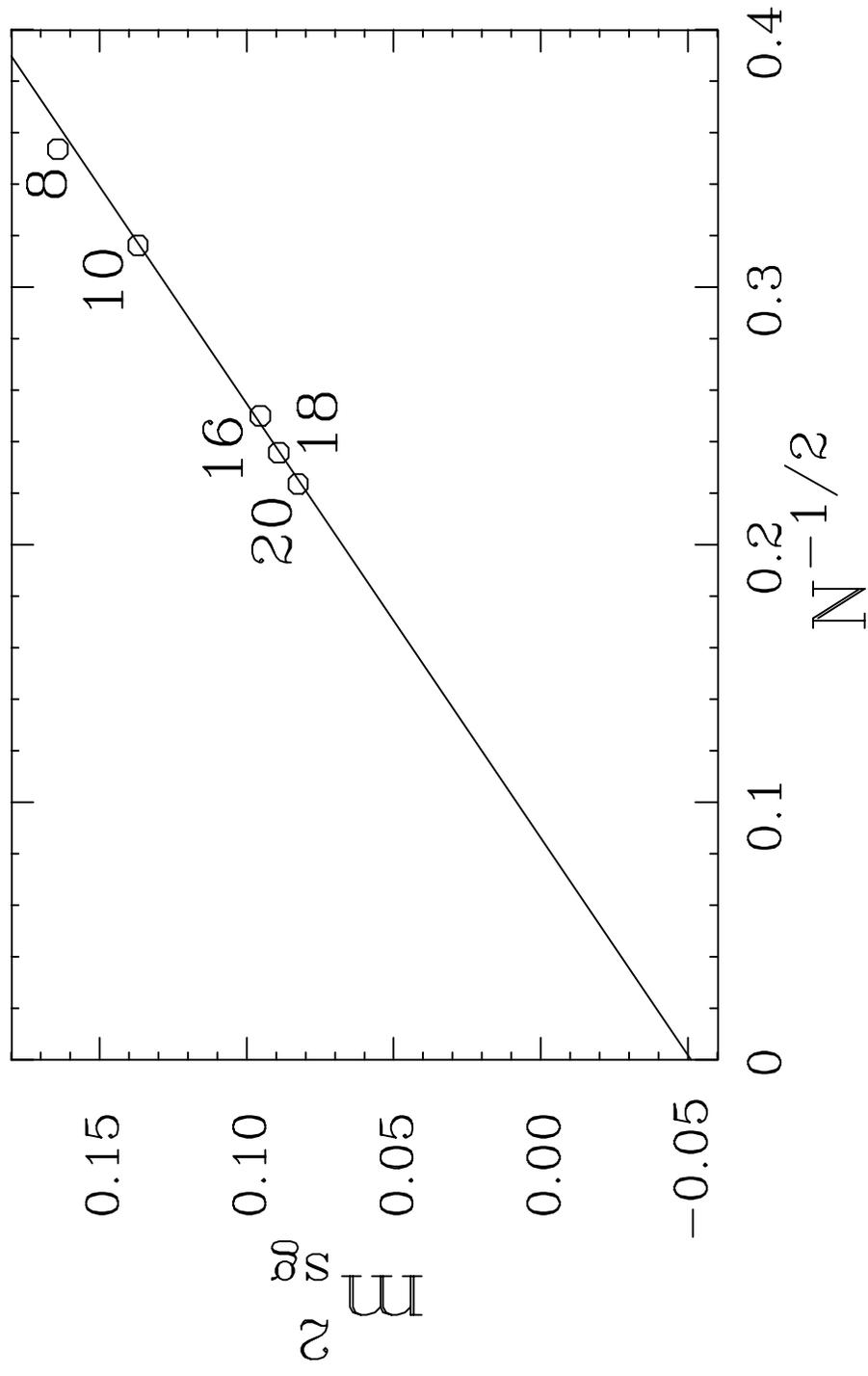
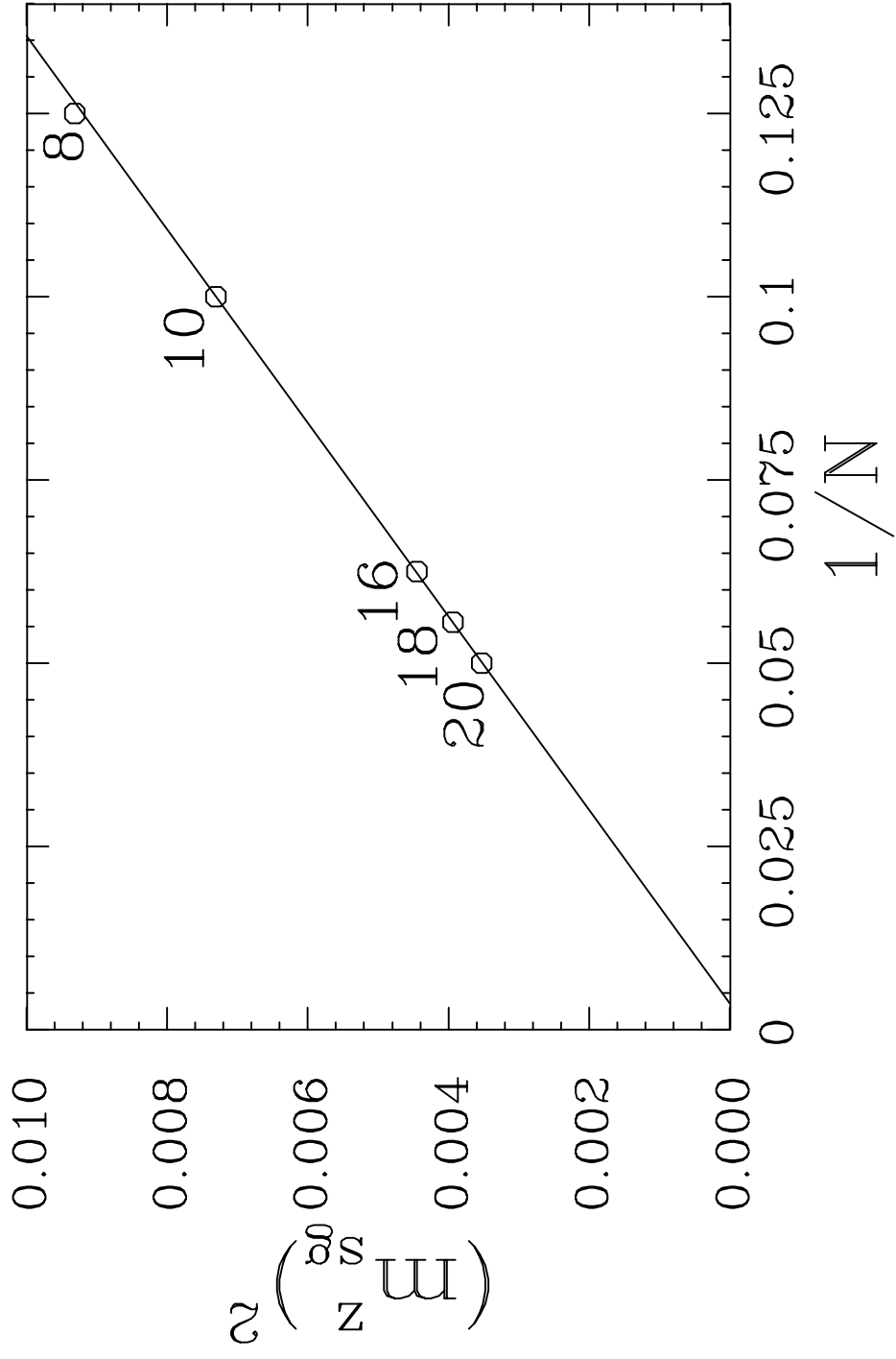
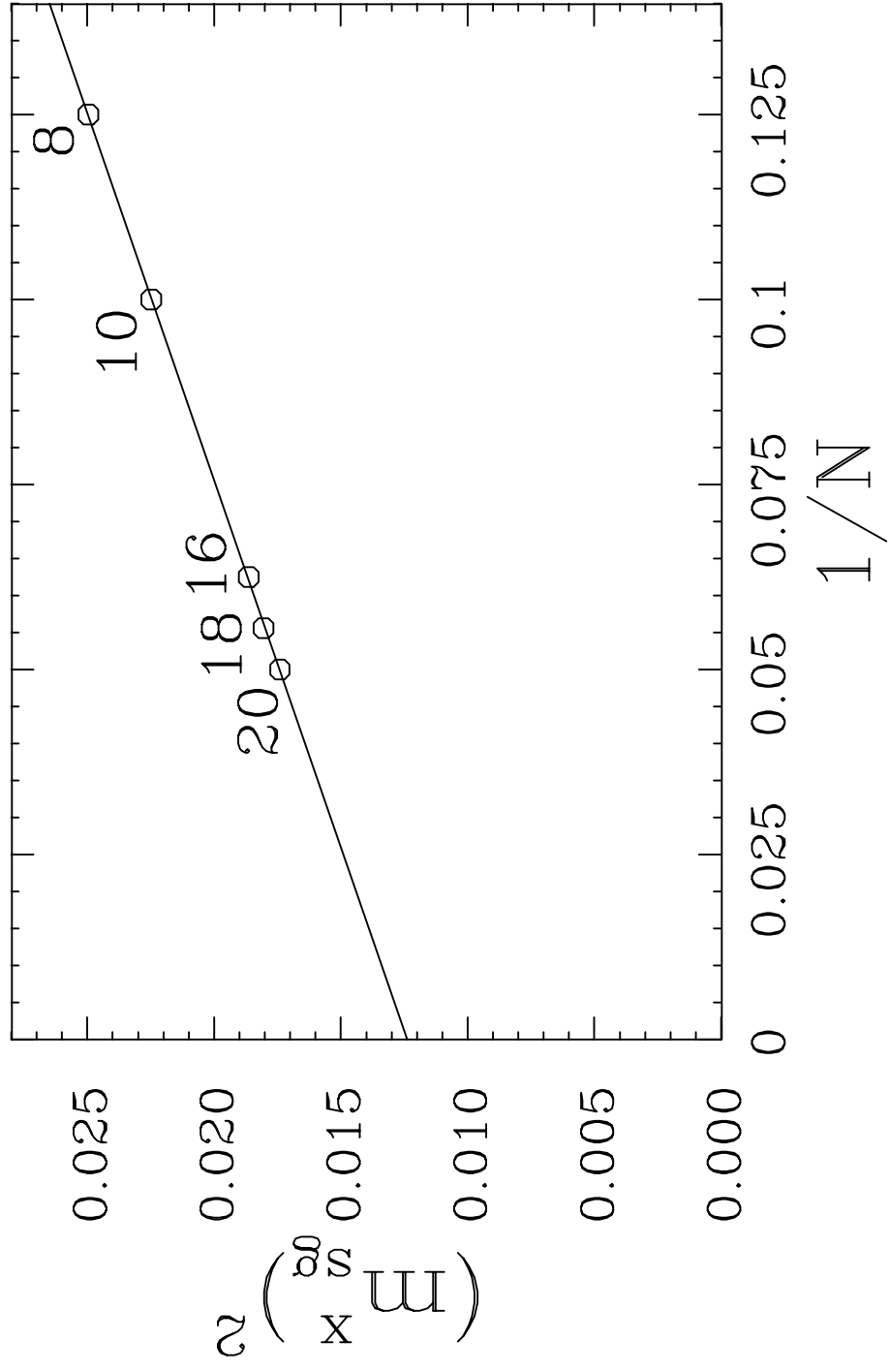
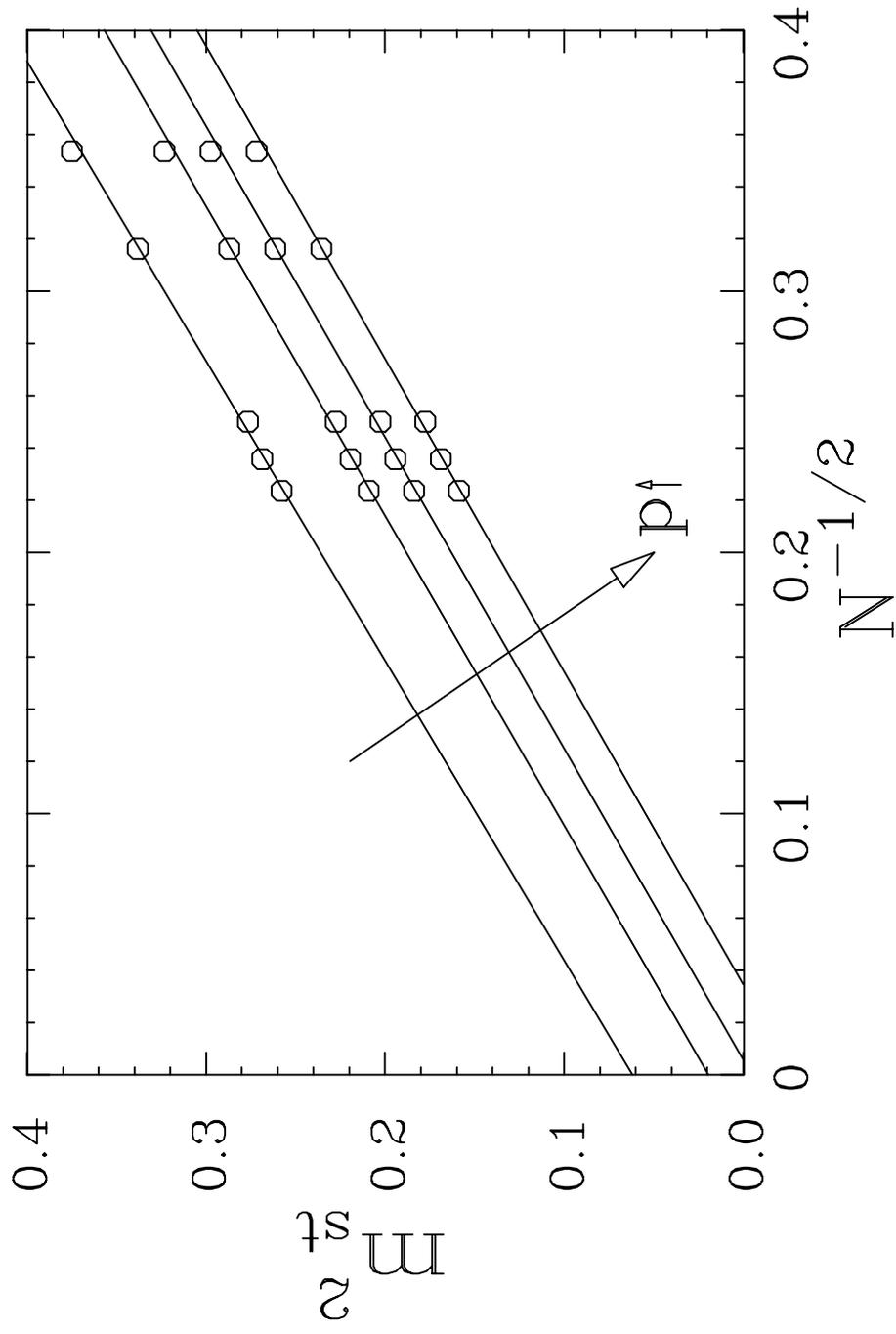
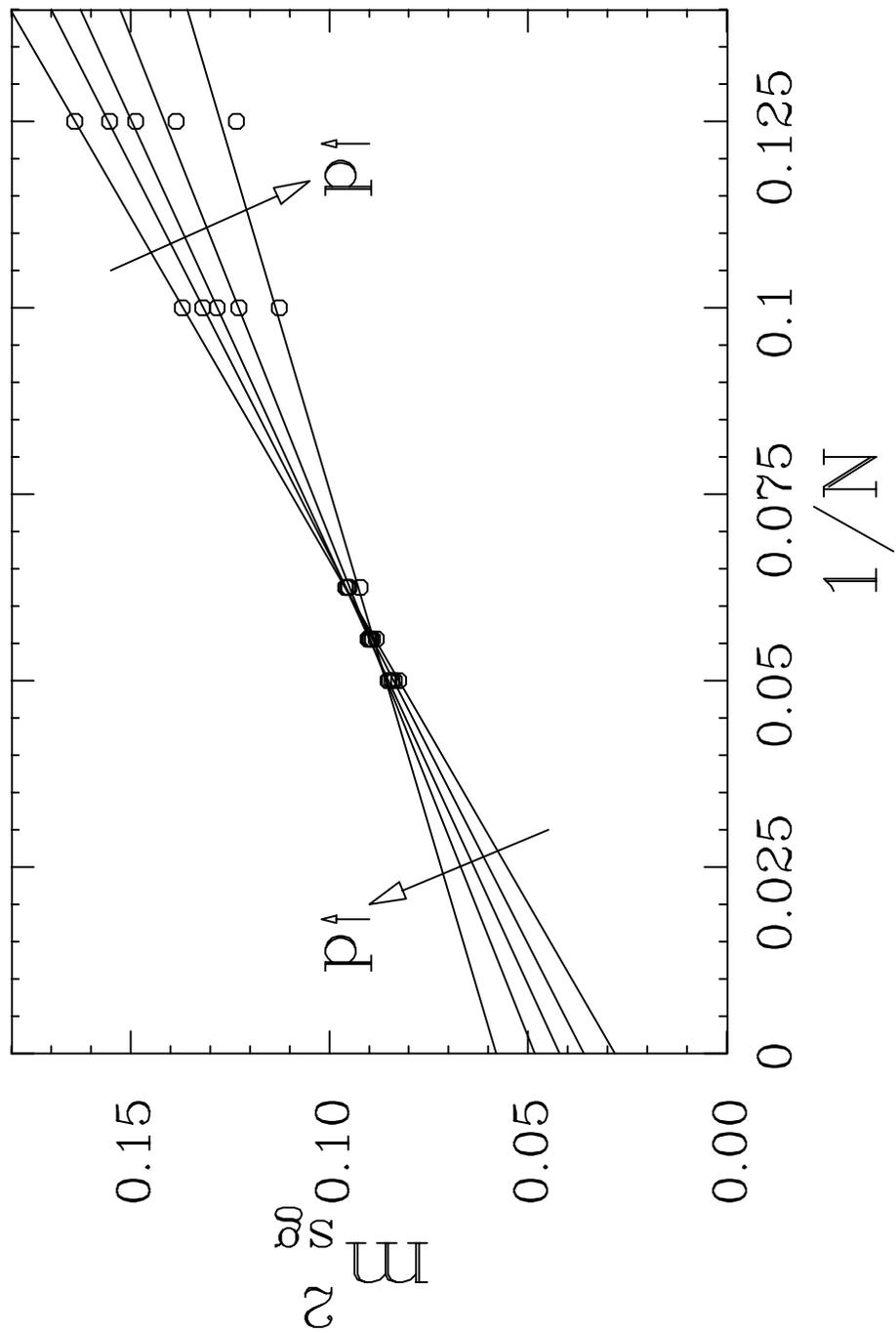

Next, we investigate the stability of the Mattis SG phase
against the coexistence of randomness and quantum fluctuation.
In Fig.\ \ref{sgfitfig}, the SG order parameter $m_{{\rm sg}}^{2}$
for the $8,10,16,18,20$-spin clusters are plotted versus $1/N$ for
$p=0.0,0.3,0.5,0.6,0.7$.
The extrapolated value of this order parameter increases monotonously,
as $p$ increases. This fact shows that this order parameter is reduced
only by quantum fluctuation, in spite of the coexistence of randomness.
Although the assumed size dependence (\ref{dissgsc}) is not based on a
strict theoretical background, such monotony of the estimate is
independent of scaling forms. Therefore, as long as the N\'eel
order exists at $p=0$, the above result will not be changed.
\section{Summary and Discussions}
The present numerical study suggests that the ground-state phase diagram
of the two-dimensional $S=1/2$ asymmetric Heisenberg Mattis model
is equivalent to that of the two-dimensional Ising Mattis model.
Namely, the values of the antiferromagnetic critical concentration
$p_{{\rm c}}^{{\rm AF}}$ are consistent within the fitting error in
these two models, and the SG order parameter in the Heisenberg Mattis
model increases monotonously as $p$ increases. These facts show that
magnetic orderings in the two-dimensional Heisenberg Mattis model is mainly
controlled by randomness, and quantum fluctuation only reduces the size of
spins. If we can identify quantum fluctuation with thermal fluctuation,
this result can be understood clearly, because the N\'eel-Mattis SG
phase boundary is independent of temperature (i.e.\ vertical)
in the Ising Mattis model. If this picture can be justified, similar
quantum-classical relationship is expected in the two-dimensional
$\pm J$ SG model, because the boundary of the N\'eel phase has been
proved vertical\cite{vertsg,vertgauge} in the classical case.

The present study also shows that ground-state properties of random
quantum spin systems can be evaluated from relatively small clusters
in the antiferromagnetic region. This fact is quite encouraging for
further studies on ground-state properties of quantum spin glasses.
Some studies in this direction\cite{site1,site2} are now in progress.
\section*{Acknowledgements}
The present author would like to thank Dr.~Y.~Ozeki for stimulating
discussions and helpful suggestions. He also thanks Prof.~H.~Nishimori
for useful comments. He is grateful for the financial support of the
Japan Society for the Promotion of Science for Japanese Junior Scientists.
This work was supported by Grant-in-Aid for Encouragement of Young
Scientists, from the Ministry of Education, Science and Culture.
The computer programs are based on the Hamiltonian-diagonalization
subroutine package ``TITPACK Ver.\ 2" developed by Professor H.~Nishimori,
its improved subroutine developed by Dr.~T.~Nakamura, and its modified
version ``KOBEPACK Ver.\ 1.0" developed by Professor T.~Tonegawa, Professor
M.~Kaburagi and Dr.~T.~Nishino. The numerical calculations were performed
on HITAC S3800/480 at the Computer Center, University of Tokyo and on the
Hewlett-Packard Apollo 735 workstation at Suzuki laboratory, University of
Tokyo. Finally, the present author thanks the members of Suzuki laboratory,
for kind permission of the usage of their computer resources.
\vfil
\pagebreak
\par

\vfil
\pagebreak
\par
\noindent
\section*{Figure Captions}
\noindent
Fig.\ 1:
Phase diagram of the Ising Mattis model.
$T_{{\rm c}}$ denotes the critical point of the corresponding
pure Ising model, and $p_{{\rm c}}$ is related with $T_{{\rm c}}$.
$[\braket{S^{z}}]_{{\rm r}}=0$ and $[\braket{S^{z}}^{2}]_{{\rm r}}=0$
in the para phase,
$[\braket{S^{z}}]_{{\rm r}}\neq 0$ and $[\braket{S^{z}}^{2}]_{{\rm r}}\neq 0$
in the ferromagnetic phase,
and $[\braket{S^{z}}]_{{\rm r}}=0$ and $[\braket{S^{z}}^{2}]_{{\rm r}}\neq 0$
in the Mattis spin glass phase.
\bigskip
\par
\noindent
Fig.\ 2:
The spin-glass order parameter of the two-dimensional $S=1/2$
antiferromagnetic Heisenberg model on a square lattice is plotted
versus $N^{-1/2}$. The number represents the size of clusters,
and the straight line is drawn by the least-squares fitting.
\bigskip
\par
\noindent
Fig.\ 3:
The spin-glass order parameter ($z$-component) of the
two-dimensional $S=1/2$ XY model on a square lattice is
plotted versus $1/N$. The number represents the size of clusters,
and the straight line is drawn by the least-squares fitting.
\bigskip
\par
\noindent
Fig.\ 4:
The spin-glass order parameter ($x$-component) of the
two-dimensional $S=1/2$ XY model on a square lattice is
plotted versus $1/N$. The number represents the size of clusters,
and the straight line is drawn by the least-squares fitting.
\bigskip
\par
\noindent
Fig.\ 5:
The N\'eel order parameter of the two-dimensional $S=1/2$ Heisenberg
Mattis model is plotted versus $N^{-1/2}$ for $p=0.0,0.1,0.15,0.2$.
The straight lines are drawn by the least-squares fitting.
\bigskip
\par
\noindent
Fig.\ 6:
The spin-glass order parameter of the two-dimensional $S=1/2$ Heisenberg
Mattis model is plotted versus $1/N$ for $p=0.0,0.3,0.5,0.6,0.7$.
The straight lines are drawn by the least-squares fitting.
\vfil
\pagebreak
\par
\noindent
\begin{figure}[h]
\caption{}
\label{asmatfig}
\end{figure}
\begin{figure}[h]
\caption{}
\label{sghffig}
\end{figure}
\begin{figure}[h]
\caption{}
\label{xysgzfig}
\end{figure}
\begin{figure}[h]
\caption{}
\label{xysgxfig}
\end{figure}
\begin{figure}[h]
\caption{}
\label{magfitfig}
\end{figure}
\end{document}